# PHYSICAL PROPERTIES OF STRANGELETS

Jes Madsen

*Institute of Physics and Astronomy, University of Aarhus, DK-8000 Århus C, Denmark, jesm@dfi.aau.dk*

Low-baryon number lumps of strange quark matter, strangelets, are presently searched for in ultra-relativistic heavy-ion experiments at CERN and Brookhaven. This paper gives an overview of the physical properties of strangelets with emphasis on experimental signatures such as mass and charge. Direct solutions of the Dirac equation with MIT bag model boundary conditions are applied and compared to calculations based on a smoothed density of states, as well as to simple mass-formulae based on the MIT bag. Most strangelet properties can be understood within an approximation including bulk, surface and curvature energy contributions.

## INTRODUCTION

A range of strong interaction parameters exists for which bulk strange quark matter (SQM) (1,2) is absolutely stable relative to nuclear matter ($E/A <$ 930MeV), or metastable on a weak interaction time-scale. The reason is that a third Fermi-sea, that of strange quarks, makes it possible to lower the energy relative to systems composed of up and down quarks only. To create a more stable system, the energy gained must first compensate for the current mass of the strange quark, but since the typical Fermi-energies involved are $m_{\text{nucleon}}/3 \approx 310$MeV, this can be the case if the strange quark is not too massive (the current mass of the strange quark is believed to be 100–300 MeV).

(Meta)stability of bulk SQM has far-reaching consequences for neutron star physics, and perhaps cosmology (for a recent review with emphasis on the astrophysics applications, see (3); for a general overview on SQM, including an extensive list of references up to mid-1991, see (4); for updates on some SQM-related issues, see (5); an overview of strangelet experiments is given in (6)). Attempts to produce SQM in the laboratory necessarily focus on low-baryon number objects (strangelets). These are in general significantly destabilized relative to bulk SQM due to finite size effects, so even if strangelets do not appear in the ultra-relativistic heavy-ion experiments at CERN and Brookhaven, this does not rule out the (meta)stability of SQM in bulk. Fortunately most of the experiments are sensitive to strangelets even if they decay on a weak interaction time-scale, and the experimental signature (first of all a very low charge-to-mass ratio) is quite unambiguous (though confusion with strange hadronic matter, discussed by Dover elsewhere in this volume, is a possibility), so the next few years will very likely either prove the existence of strangelets or put significant constraints on strong interaction model parameters.

In the following I will introduce a physical framework for understanding the main properties of strangelets. Starting with a brief discussion of the MIT bag model and shell model calculations for strangelets, I will go on to describe a simple strangelet mass-formula, which I later derive explicitly from the MIT bag model. This allows a comparison of strangelet properties as a function of parameters. Finally I stress the consequences of the results in relation to experimental efforts. The main message is that some strangelet properties (like low charge-to-mass ratio) are quite robust, but that for instance the masses (and therefore also the stability properties) are very uncertain due to the freedom in choosing parameters. Some parts of the manuscript are updated versions of another recent review (7).

## STRANGE QUARK MATTER WITHIN THE MIT BAG MODEL

In the simplest version of the MIT bag model (8) noninteracting quarks are confined in a spherical cavity of radius $R$. They satisfy the free Dirac equation inside the cavity and obey a boundary condition at the surface, which corresponds to no quark current flow across the surface. The bag itself has an energy of $BV$, where the bag constant, $B$, is a measure of the false vacuum energy that confines the quarks ($B$ thus in practice acts like an external pressure on the bag). In the simplest version the energy (mass) of the system is given by the sum of the bag energy and the energies of individual quarks,

$$E = \sum_{i=u,d,s} \sum_{\kappa} N_{\kappa,i}(m_i^2 + k_{\kappa,i}^2)^{1/2} + B 4\pi R^3/3. \qquad (1)$$

Here the momentum $k_{\kappa,i}$ is found from the Dirac equation. For states with quantum numbers $(j,l)$, $\kappa = \pm(j+\frac{1}{2})$ for $l = j \pm \frac{1}{2}$. For a given quark flavor each level has a degeneracy of $N_{\kappa,i} = 3(2j+1)$ (the factor 3 from color degrees of freedom). For example, the $1S_{1/2}$ ground-state ($j = 1/2$, $l = 0$, $\kappa = -1$) has a degeneracy of 6.

The level filling scheme is somewhat cumbersome. To find the ground state strangelet for fixed bag constant, quark-masses, and baryon number, one must fill up the lowest energy levels for a choice of radius; then vary the radius until a minimum energy is found ($\partial E/\partial R = 0$). Since levels cross, the order of levels is changing as a function of $R$.

Results of such calculations are shown in Figure 1. Notice that the energy per baryon smoothly approaches a bulk limit for $A \to \infty$, whereas the energy grows significantly for low $A$. For low $s$-quark mass a very significant shell appears for $A = 6$ (3 colors and 2 spin orientations per flavor), and less conspicuous ones for a range of higher baryon numbers. As $m_s$ increases it becomes more and more favorable to use $u$ and $d$ rather than $s$-quarks, and the "magic numbers" change; for instance the first closed shell is seen for $A = 4$ rather than 6.

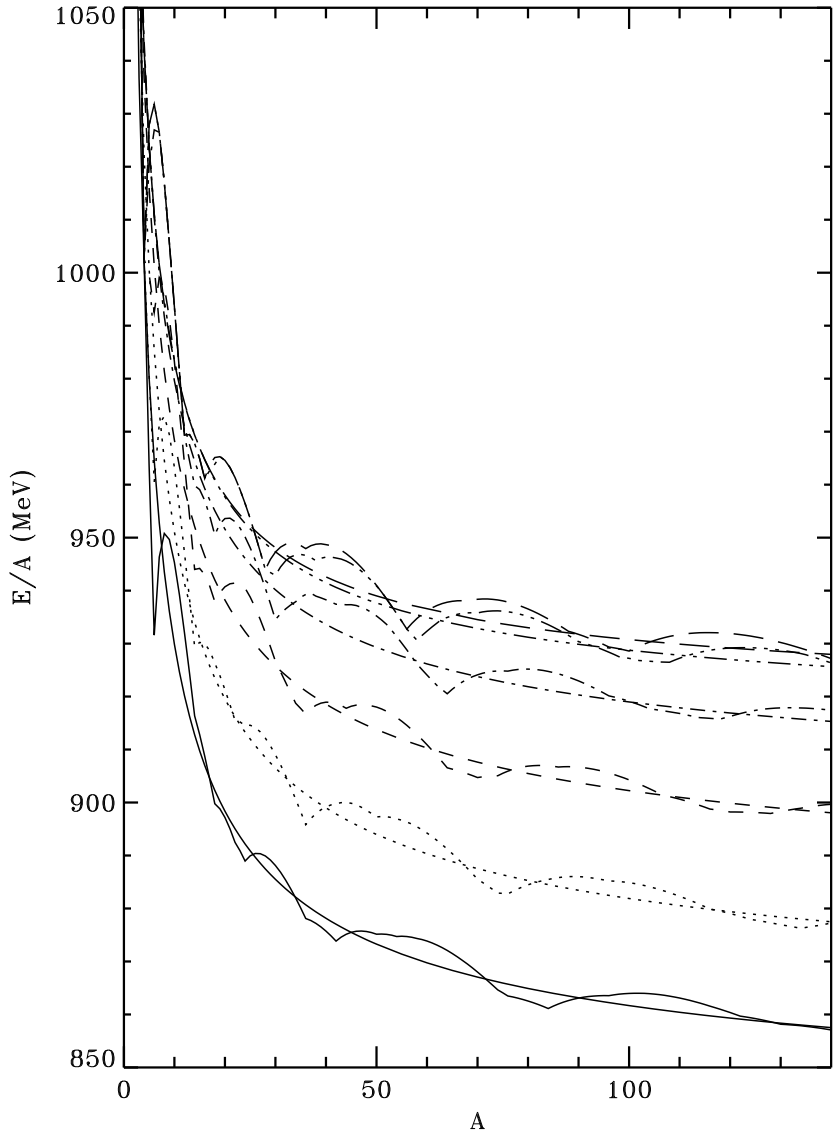

**FIG. 1.** Energy per baryon as a function of baryon number for $B^{1/4} = 145\,\text{MeV}$ and $s$-quark mass from 50–300 MeV in steps of 50 MeV (higher curves have higher $m_s$). Spiky curves are based on direct mode-filling calculations, smooth curves on the smoothed density of states from the multiple reflection expansion. The smooth curves are thus not fits to the shell model calculations, but derived from a smoothing procedure *within* the MIT bag model.

Equation (1) can be modified by inclusion of Coulomb energy, zero-point fluctuation energy, and gluon exchange energy. The zero-point energy is normally included as a phenomenological term of the form $-Z_0/R$, where fits to light hadron spectra (8) indicate the choice $Z_0 = 1.84$ for $B_{145}^{1/4} \equiv B^{1/4}/145\text{MeV} = 1$. In the MIT bag model the zero point energy is to some extent a fudge factor adjusted to fit the spectra. Roughly half of this phenomenological term has a physical basis in a center-of-mass motion, which can be subtracted in a more direct fashion. The proper choice of $Z_0$ (or even the functional form of the zero-point energy) is not straightforward. As discussed by Farhi and Jaffe (9) the value is intimately coupled to $B$ and $m_s$, as well as to the strong fine-structure constant, $\alpha_s$, and it is not obvious that values deduced from bag model fits to ordinary hadrons are to be preferred. For reasonable parameter values one sees a significant effect of the zero-point energy for $A < 10$, but the term quickly becomes negligible for increasing $A$, where its contribution to the energy per baryon goes like $A^{-4/3}$. Because of the great uncertainty, no zero-point energy is included in the numerical results presented in this overview.

Bulk systems are described in terms of 3 parameters: The bag constant, $B$; the strange quark mass, $m_s$ (up and down quarks are normally assumed to be massless); and $\alpha_s$, the strong fine-structure constant, which describes gluon exchange interactions. For most purposes a non-zero $\alpha_s$ can be "absorbed" in a reduction of $B$, so in the following I shall concentrate on $\alpha_s = 0$. A lower limit on $B$ ($B^{1/4} > 145\text{MeV}$) can be inferred from the stability of ordinary nuclei relative to up-down quark matter (ordinary nuclei do not spontaneously decay into strangelets since this would require a high order weak interaction to make sufficient numbers of strange quarks). A bag constant smaller than $(164\text{MeV})^4$ permits stable bulk SQM for sufficiently low $m_s$, whereas the metastability window relative to a gas of $\Lambda$'s goes to $(195\text{MeV})^4$.

## STRANGELET MASS-FORMULAE

Ordinary nuclei are often discussed in terms of mass-formulae, where the energy per baryon is a sum of constituent masses, Coulomb energy, surface energy, symmetry energy, and perhaps other, less important terms. In a rather similar fashion, the energy per baryon for strangelets, $E/A \equiv \epsilon$, can be written as a bulk term, $\epsilon^0$, plus a number of finite size corrections,

$$\epsilon = \epsilon^0 + \epsilon_{\text{Coulomb}} + \epsilon_{\text{surface}} + \epsilon_{\text{curvature}} + \epsilon_{\text{zero}}... \tag{2}$$

In contrast to nuclei, $\epsilon_{\text{Coulomb}}$ is negligible for most strangelets due to the cancellation of $u$, $d$, and $s$ quark charges. The Coulomb-term is important for determining the ground state composition of strangelets, but the energy itself is negligible in the mass-formula. The surface tension is important as in nuclear physics, but more important for most choices of bag model parameters is the curvature energy, which plays a negligible role in nuclei. Both terms are a consequence of the bag model boundary conditions, which result in a depletion of massive $s$-quarks near the surface (this is further discussed below).

In a first approximation strangelets have constant density (slightly higher than the nuclear matter density) for fixed bag parameters, so that baryon number $A \propto R^3$, $E_{\mathrm{surf}} \propto R^2$, and $E_{\mathrm{curv}} \propto R$, where $R$ is the strangelet radius. In this approximation (where we neglect $\epsilon_{\mathrm{zero}}$ for the time being)

$$\epsilon \approx \epsilon^0 + c_{\mathrm{surf}} A^{-1/3} + c_{\mathrm{curv}} A^{-2/3}, \tag{3}$$

with $c_{\mathrm{surf}} \approx 100 \mathrm{MeV}$ and $c_{\mathrm{curv}} \approx 300 \mathrm{MeV}$ as typical values. (The coefficients will be derived in the following subsections).

Stability relative to a gas of neutrons requires $\epsilon < m_n$, which may be written as $A > A_{\mathrm{min}}^{\mathrm{abs}}$, where

$$A_{\mathrm{min}}^{\mathrm{abs}} = \left( \frac{c_{\mathrm{surf}} + [c_{\mathrm{surf}}^2 + 4 c_{\mathrm{curv}}(m_n - \epsilon^0)]^{1/2}}{2(m_n - \epsilon^0)} \right)^3. \tag{4}$$

Stability at baryon number 30 requires a bulk binding energy in excess of 65 MeV, which is barely within reach for $m_s > 100 \mathrm{MeV}$ if, at the same time, $ud$-quark matter shall be unstable. The proposed cosmic ray strangelet-candidates with baryon number 370 (10) would for stability require a bulk binding energy per baryon exceeding 20 MeV to overcome the combined curvature and surface energies. Absolute stability relative to a gas of $^{56}$Fe corresponds to furthermore using 930 MeV instead of $m_n$, whereas stability relative to a gas of $\Lambda$-particles (the ultimate limit for formation of short-lived strangelets) would correspond to substitution of $m_\Lambda = 1116 \mathrm{MeV}$.

One can also calculate the minimum baryon number for which long-lived metastability with respect to neutron emission is possible. This requires $dE_{\mathrm{curv}}/dA + dE_{\mathrm{surf}}/dA < m_n - \epsilon^0$, or

$$A_{\mathrm{min}}^{\mathrm{meta}} = \left( \frac{c_{\mathrm{surf}} + [c_{\mathrm{surf}}^2 + 3 c_{\mathrm{curv}}(m_n - \epsilon^0)]^{1/2}}{3(m_n - \epsilon^0)} \right)^3. \tag{5}$$

To have $A_{\mathrm{min}}^{\mathrm{meta}} < 30$ requires $m_n - \epsilon^0 > 30 \mathrm{MeV}$, which is possible, but only for a narrow range of parameters.

However, it is important to notice, that shell effects can have a stabilizing influence. As stressed by Gilson and Jaffe (11) the fact that the slope of $E/A$ versus $A$ becomes very steep near magic numbers can lead to strangelets that are metastable (stable against single baryon emission) even for $\epsilon^0 > 930 \mathrm{MeV}$. Also, the time-scales for energetically allowed decays have not been calculated. Pauli-blocking is known to delay weak quark conversion in strangelets, and this will probably have a significant influence on the lifetimes.

### Smoothed density of states

I will now show how the strangelet mass-formula can be explicitly derived within the MIT bag model.

Strangelets with $A \ll 10^7$ are smaller than the electron Compton wavelength, and electrons are therefore mainly localized outside the quark phase (9), which means that they can safely be neglected in mass calculations. Strangelets therefore do not obey a requirement of local charge neutrality, as in the case of SQM in bulk. This leads to a small Coulomb energy, which is rather negligible for the mass-formula (less than a few MeV per baryon), but which is decisive for the quark composition and therefore the charge-to-mass ratio, $Z/A$, of the strangelet. A characteristic of strangelets, which is perhaps the best experimental signature, is that this ratio is very small compared to ordinary nuclei. Indeed, for $m_s = 0$ the ground state strangelet has equal numbers of all three quark flavors and is therefore charge neutral. Whereas Coulomb effects have been consistently included (12), I will leave out those terms in the equations below.

In the ideal Fermi-gas approximation the energy of a system composed of quark flavors $i$ is given by

$$E = \sum_i (\Omega_i + N_i \mu_i) + BV \qquad (6)$$

Here $\Omega_i$, $N_i$ and $\mu_i$ denote thermodynamic potentials, total number of quarks, and chemical potentials, respectively. $B$ is the bag constant, $V$ is the bag volume.

The proper calculation of the energy by an explicit solution of the Dirac equation with MIT boundary conditions was described earlier. A physically more illuminating approach, first applied to strangelets (without the curvature term, the significance of which was realized in (13)) by Farhi & Jaffe (9) and Berger & Jaffe (14), is to use a smoothed density of states rather than the explicit mode-filling scheme. In the multiple reflection expansion framework of Balian and Bloch (15), such a smoothed density of states can be written as

$$\frac{dN_i}{dk} = 6\left\{\frac{k^2 V}{2\pi^2} + f_S\left(\frac{m_i}{k}\right)kS + f_C\left(\frac{m_i}{k}\right)C + ....\right\}, \qquad (7)$$

where the area $S = \oint dS = 4\pi R^2$ for a sphere, and curvature $C = \oint \left(\frac{1}{R_1} + \frac{1}{R_2}\right)dS = 8\pi R$ for a sphere. Curvature radii are denoted $R_1$ and $R_2$. For a spherical system $R_1 = R_2 = R$. The functions $f_S$ and $f_C$ are given below.

In terms of volume-, surface-, and curvature-densities, $n_{i,V}$, $n_{i,S}$, and $n_{i,C}$, the number of quarks of flavor $i$ is

$$N_i = \int_0^{k_{Fi}} \frac{dN_i}{dk} dk = n_{i,V} V + n_{i,S} S + n_{i,C} C, \qquad (8)$$

with Fermi momentum $k_{Fi} = (\mu_i^2 - m_i^2)^{1/2} = \mu_i(1 - \lambda_i^2)^{1/2}$; $\lambda_i \equiv m_i/\mu_i$.

The corresponding thermodynamic potentials are related by

$$\Omega_i = \Omega_{i,V} V + \Omega_{i,S} S + \Omega_{i,C} C, \qquad (9)$$

where $\partial \Omega_i/\partial \mu_i = -N_i$, and $\partial \Omega_{i,j}/\partial \mu_i = -n_{i,j}$. The universal volume terms are given by

$$\Omega_{i,V} = -\frac{\mu_i^4}{4\pi^2}\left((1-\lambda_i^2)^{1/2}(1-\frac{5}{2}\lambda_i^2) + \frac{3}{2}\lambda_i^4 \ln\frac{1+(1-\lambda_i^2)^{1/2}}{\lambda_i}\right), \qquad (10)$$

$$n_{i,V} = \frac{\mu_i^3}{\pi^2}(1-\lambda_i^2)^{3/2}. \qquad (11)$$

The surface and curvature contributions are derived from (14)

$$f_S\left(\frac{m}{k}\right) = -\frac{1}{8\pi}\left\{1 - \frac{2}{\pi}\tan^{-1}\frac{k}{m}\right\}, \qquad (12)$$

and (16)

$$f_C\left(\frac{m}{k}\right) = \frac{1}{12\pi^2}\left\{1 - \frac{3}{2}\frac{k}{m}\left(\frac{\pi}{2} - \tan^{-1}\frac{k}{m}\right)\right\}. \qquad (13)$$

The explicit forms of $\Omega_{i,S}$ and $n_{i,S}$ were first given in (14), and $\Omega_{i,C}$ and $n_{i,C}$ in (16). The cumbersome formulae will not be quoted here, but their properties are shown in Figure 2. Note in particular that whereas $\Omega_{i,S}$ and $\Omega_{i,C} \to 0$ for $m \to \mu$ (no quarks of flavor $i$ are present for $m > \mu$), and $\Omega_{i,S} \to 0$ also for $m \to 0$, the curvature energy given by $\Omega_{i,C}$ is non-zero for massless quarks. This is very important because it means, that the nearly massless $u$ and $d$ quarks, while not contributing to the surface tension, give a very large contribution to the curvature energy. This effect was forgotten in strangelet studies prior to 1993, leading to mass-formulae that were too optimistic (significantly underestimating the energy per baryon) at the low baryon numbers accessible in ultra-relativistic heavy-ion colliders (13).

An equilibrium strangelet is found by minimizing the energy, i.e. from

$$0 = dE = \sum_i (\Omega_{i,V} dV + \Omega_{i,S} dS + \Omega_{i,C} dC + \mu_i dN_i) + B dV. \qquad (14)$$

Minimizing at fixed $N_i$ for a sphere gives the pressure equilibrium constraint

$$B = -\sum_i \Omega_{i,V} - \frac{2}{R}\sum_i \Omega_{i,S} - \frac{2}{R^2}\sum_i \Omega_{i,C}. \qquad (15)$$

The "optimal" composition is found by minimizing the energy with respect to $N_i$ at fixed $A$ and radius, giving

$$0 = \sum_i \mu_i dN_i. \qquad (16)$$

Together with Eq. (6) these constraints give the properties of a spherical quark lump in its ground state.

The solution is compared with the shell model calculations in Figure 1. The agreement is remarkably good, indicating that all the important physics (apart from the shells) can be understood in terms of the surface and curvature contributions.

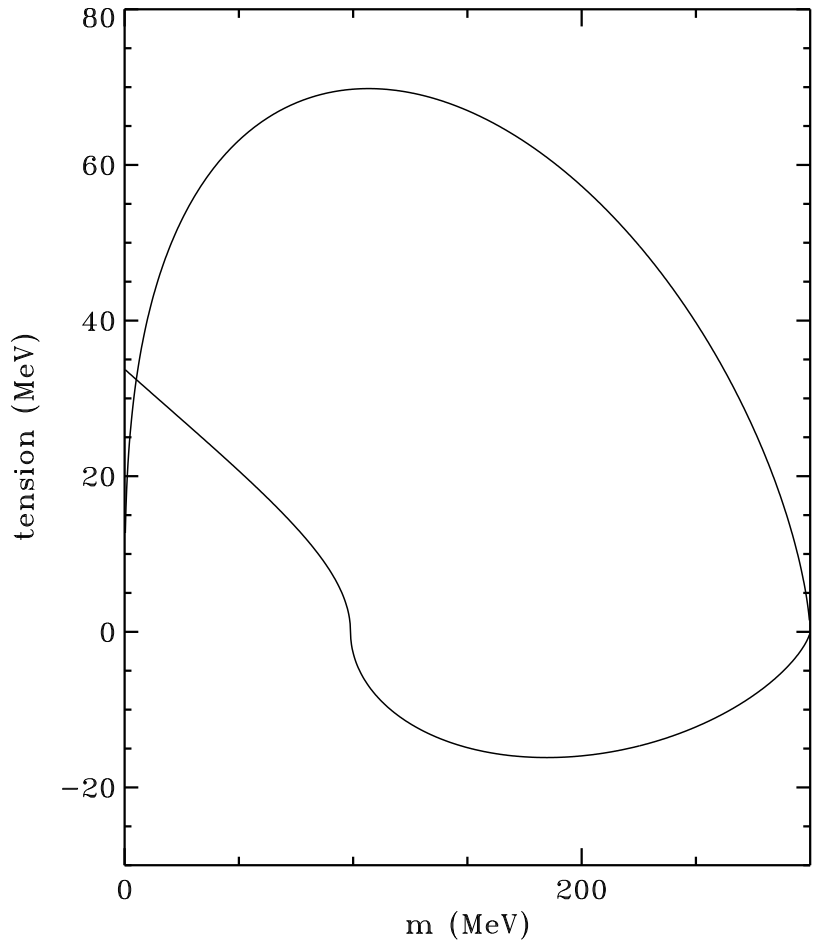

**FIG. 2.** Surface tension and curvature coefficient as a function of quark mass for quark chemical potential of 300 MeV. The upper curve is $\Omega_{i,S}^{1/3}$, while the lower curve is $|\Omega_{i,C}|^{1/2}$ multiplied by the sign of the curvature term. Numbers on both axes scale in proportion to $\mu$. Note the significant curvature contribution for massless quarks.

### Bulk approximation

Simpler, approximate mass-formulae can be derived by using a bulk approximation to the chemical potentials. This was the approach taken in the first detailed investigation of the strangelet mass-formula within the MIT bag model performed by Berger and Jaffe (14). They included Coulomb corrections and surface tension effects stemming from the depletion in the surface density of states due to the mass of the strange quark. Both effects were treated as perturbations added to a bulk solution with the surface contribution derived from Eq. (12).

While showing some of the important physics, this approach (apart from Coulomb corrections) predicted constant $E/A$ versus $A$ for $m_s \to 0$ and $m_s \to \mu_s$, in striking contrast to the shell model results, because the very important curvature term was not included. Below I discuss the bulk approximation with curvature.

Including bulk terms only, the energy minimization, Eq. (15) (with $\lambda \equiv \lambda_s$, and superscript $^0$ denoting bulk values), changes to

$$B = -\sum_i \Omega^0_{i,V}$$
$$= \sum_{i=u,d} \frac{(\mu^0_i)^4}{4\pi^2} + \frac{(\mu^0_s)^4}{4\pi^2}\left[(1-\lambda^2)^{1/2}(1-\frac{5}{2}\lambda^2)\right.$$
$$\left. +\frac{3}{2}\lambda^4 \ln \frac{1+(1-\lambda^2)^{1/2}}{\lambda}\right], \qquad (17)$$

and the baryon number density is now given by

$$n^0_A = \frac{1}{3}\left[\sum_{i=u,d} \frac{(\mu^0_i)^3}{\pi^2} + \frac{(\mu^0_s)^3}{\pi^2}(1-\lambda^2)^{3/2}\right]. \qquad (18)$$

A bulk radius per baryon is defined by

$$R^0 \equiv (3/4\pi n^0_A)^{1/3}. \qquad (19)$$

In bulk equilibrium the chemical potentials of the three quark flavors are equal, $\mu^0_u = \mu^0_d = \mu^0_s = \epsilon^0/3$, where $\epsilon^0$ is the bulk energy per baryon. One may approximate the energy per baryon of small strangelets as a sum of bulk, surface and curvature terms, using the chemical potential calculated in bulk:

$$\epsilon \approx \epsilon^0 + A^{-1}\sum_i \Omega^0_{i,S}S^0 + A^{-1}\sum_i \Omega^0_{i,C}C^0, \qquad (20)$$

where $S^0 = 4\pi(R^0)^2 A^{2/3}$ and $C^0 = 8\pi(R^0)A^{1/3}$. Examples for $B^{1/4} = 145$MeV are (with $s$-quark mass given in parenthesis; all energies in MeV)

$$\epsilon(0) = 829 + 0A^{-1/3} + 351A^{-2/3}$$

$$\epsilon(50) = 835 + 61A^{-1/3} + 277A^{-2/3}$$
$$\epsilon(100) = 852 + 83A^{-1/3} + 241A^{-2/3}$$
$$\epsilon(150) = 874 + 77A^{-1/3} + 232A^{-2/3}$$
$$\epsilon(200) = 896 + 53A^{-1/3} + 242A^{-2/3}$$
$$\epsilon(250) = 911 + 22A^{-1/3} + 266A^{-2/3}$$
$$\epsilon(300) = 917 + 0A^{-1/3} + 295A^{-2/3}. \tag{21}$$

The bulk approximations above generally undershoot the correct solution with properly smoothed density of states (smooth curves in Figure 1) by 2MeV for $A > 100$, 5MeV for $A \approx 50$, 10MeV for $A \approx 10$ and 20MeV for $A \approx 5$. This is because the actual chemical potentials of the quarks increase when $A$ decreases, whereas the bulk approximations use constant $\mu$. For massless $s$-quarks the expression for $\epsilon(0)$ scales simply as $B^{1/4}$. The same scaling applies for $m_s > \epsilon^0/3$, where no $s$-quarks are present; in the example above the scaling can be applied to $\epsilon(300)$. For intermediate $s$-quark masses both $\epsilon$ and $m_s$ should be multiplied by $B_{145}^{1/4}$ to scale the results. For $B^{1/4} = 165$MeV one finds $\epsilon(150) = 985 + 93A^{-1/3} + 265A^{-2/3}$; $\epsilon(250) = 1027 + 46A^{-1/3} + 284A^{-2/3}$. Coulomb effects were not included here. They have no influence for $m_s \to 0$, but change the results by a few MeV for large $m_s$.

In connection with the shell-model calculations I described the effects of a zero-point energy of the form $-Z_0/R$, and claimed that it was important only for $A < 10$. This can be understood in the bulk approximation of constant $\mu$, because the zero-point term per baryon is proportional to $A^{-4/3}$ compared to $A^{-1/3}$ and $A^{-2/3}$ for surface and curvature energies. The full term to be added to the bulk approximation expressions for a given $\epsilon^0$ is:

$$\epsilon_{\text{zero}} = -Z_0 (4/243\pi)^{1/3} \left[ 2 + [1 - (3m_s/\epsilon^0)]^{3/2} \right]^{1/3} \epsilon^0 A^{-4/3}, \tag{22}$$

typically of order $-200 Z_0 \text{MeV} A^{-4/3}$.

## EXCITED STATES

The calculations discussed so far have focussed on finding the ground state properties of strangelets, i.e. the lowest mass state for a given baryon number, $A$. Figure 3, which is based on direct mode-filling including Coulomb energy, instead fixes $A = 20$ and displays contours of equal $E/A$ for the whole range of possible strangelets with that baryon number (a total of 1891 states). What is quite interesting from an experimental point of view is the close spacing of states around that of minimum energy. More than a hundred different states are within 10MeV per baryon from the ground state. The ground state itself has slightly positive charge for all choices of parameters tried so far, but neutral as well as negatively charged states are very close in energy. Many of these states will be stable against strong decays, and since some weak decays are suppressed by Pauli-blocking, presumably many different configurations could be sufficiently long-lived to reach the detectors. This may be good from a production point of view, but also means that there is not going to be *a single* well-defined strangelet signature, but rather numerous (meta)stable states and many more possible decay modes!

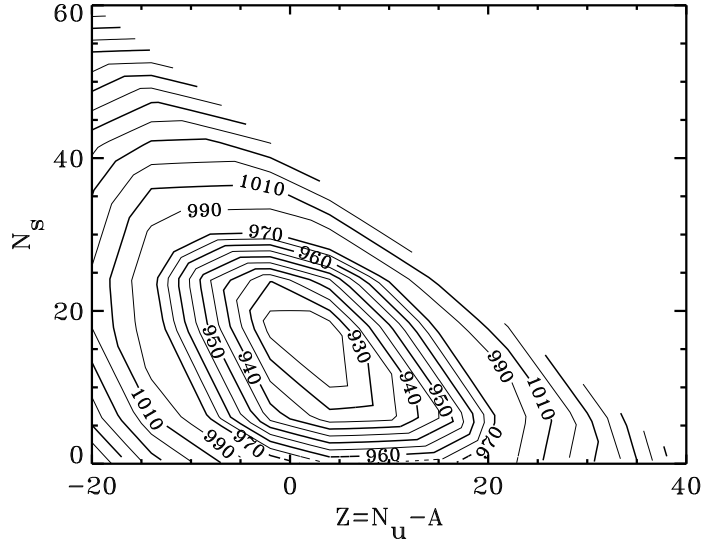

**FIG. 3.** Contours of energy per baryon in MeV for all 1891 strangelet configurations with $A = 20$, $m_s = 100$MeV, and $B_{145}^{1/4} = 1$. Contours are plotted as a function of charge ($Z = N_u - A$) and the number of strange quarks, $N_s$.

Related results are shown in Figure 4 where the mass-formula in the bulk approximation has been used to compare $E/A$ for strangelets of equal mass and charge. As is seen there, even for fixed $B$, $m_s$, $A$, and $Z$, there is a huge span in strangelet mass! High resolution mass measurements may therefore be of significant interest for comparison between theory and experiment.

## CONCLUSIONS

The physical properties of strangelets can be summarized as follows:

1) (Meta)stable strangelets are possible for a wide range of parameters.

2) Energy per baryon, charge etc are strongly parameter dependent.

3) Normally $|Z| \ll A$, and the ground state strangelets have slightly positive charges.

4) A liquid drop model (Fermi-gas model) explains the general properties. Curvature is a decisive effect.

5) Strangelets have prominent shell structure for low $A$.

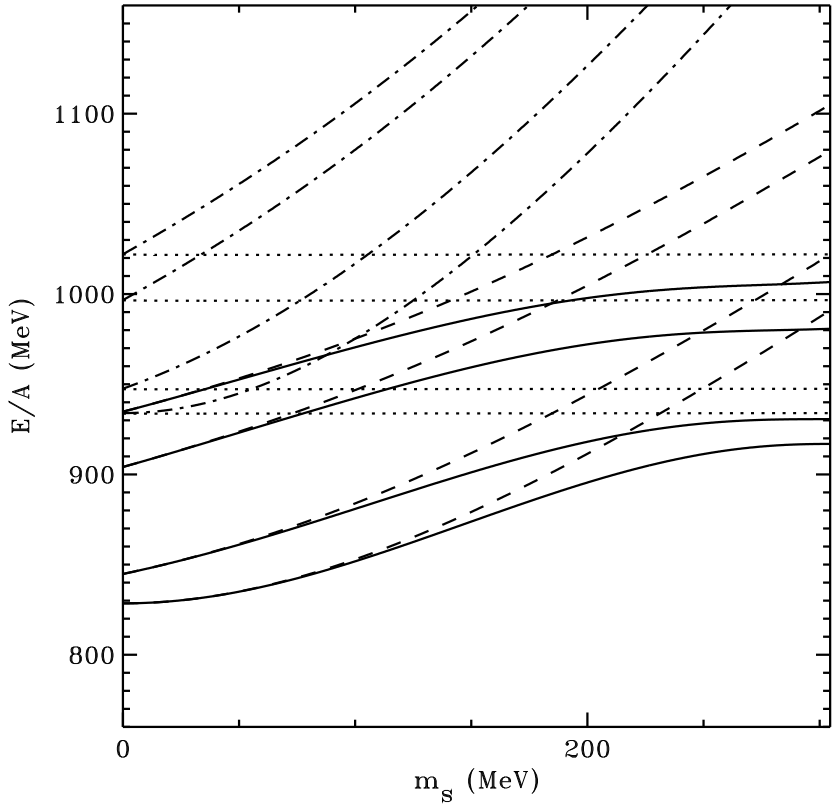

**FIG. 4.** Energy per baryon in MeV as a function of strange quark mass for $B^{1/4} = 145$ MeV. Four sets of curves are shown. For each set of curves the one with lowest $E/A$ represents bulk SQM, whereas the others from below are for baryon numbers $A = 100$, 10, and 6, respectively. Full curves show the ground state strangelets, i.e. those with the lowest energy per baryon for the given $A$. The other curves all have fixed charge, $Z = 0$, but varying strangeness content. Dotted, horizontal curves have zero strangeness ($S/A = 0$, where the strangeness $S$ is minus the number of strange quarks). Dashed curves have $S/A = -1$, and dash-dot curves have $S/A = -2$ (and no $d$-quarks). Results like these obtained within the bulk approximation to the MIT bag model can be scaled to other choices of $B$ by multiplying numbers on both axes by $B^{1/4}_{145}$. Notice the wide span of possible strangelet masses for fixed $B$, $A$, $Z$, and $m_s$, depending on the quark composition of the strangelet! The composition of experimentally accessible strangelets is again determined by the production mechanism and the decay properties of the excited states; accurate mass measurements would be of great importance.

6) Many isotopes are almost degenerate in energy.

7) Decay modes for metastable strangelets are numerous. Lifetimes are not predictable at present, but could well be larger (due to Pauli-blocking) than those of the strange hadronic matter discussed by Dover elsewhere in these Proceedings, thus providing a possible experimental distinction (also, strange hadronic matter can not be absolutely stable in contrast to strangelets).

The conclusions above are mainly based on studies within the simplest version of the MIT bag model. Effects of zero-point energy and finite $\alpha_s$ have only been included in a crude fashion, and actual QCD calculations are beyond reach. Real strangelets can have non-spherical shape. They are created at a high temperature (which tends to increase the energy per baryon). Et cetera.

Work on some of these complications is in progress, but clearly our theoretical understanding of strangelets is still in a rather primitive state. Experimental data are desperately needed before constructing even more complicated models!

## ACKNOWLEDGMENTS

This work was supported in part by the Theoretical Astrophysics Center under the Danish National Research Foundation. I appreciate discussions with Dan Jensen, Michael Christiansen, Carl Dover, Shiva Kumar, and Horst Stöcker.